# A Unified Framework for Trees, MDS and Planar Graphs

Peter J. Waddell[1], Ishita Khan[2,] Xi Tan[2], and Sunghwan Yoo[2]

pwaddell@purdue.edu
tan19@purdue.edu
yoo7@purdue.edu
khan27@purdue.edu

[1]Department of Biological Sciences, Purdue University, West Lafayette, IN 47906, U.S.A.
[2]Department of Computer Science, Purdue University, West Lafayette, IN 47906, U.S.A

Least squares trees, multidimensional scaling and Neighbor Nets are all different and popular ways of visualizing multi-dimensional data. The method of flexi-Weighted Least Squares (fWLS) is a powerful method of fitting phylogenetic trees, when the exact form of errors is unknown. Here, both polynomial and exponential weights are used to model errors. The exact same models are implemented for Multi-Dimensional Scaling to yield flexi-Weighted MDS, including as special cases methods such as the Sammon Stress function. Here we apply all these methods to population genetic data looking at the relationships of "Abrahams Children" encompassing Arabs and now widely dispersed populations of Jews, in relation to an African outgroup and a variety of European populations. Trees, MDS and Neighbor Nets of this data are compared within a common likelihood framework and the strengths and weaknesses of each method are explored. Because the errors in this type of data can be complex, for example, due to unexpected genetic transfer, we use a residual resampling method to assess the robustness of trees and the Neighbor Net. Despite the Neighbor Net fitting best by all criteria except BIC, its structure is ill defined following residual resampling. In contrast, fWLS trees are favored by BIC and retain considerable strong internal structure following residual resampling. This structure clearly separates various European and Middle Eastern populations, yet it is clear all of the models have errors much larger than expected by sampling variance alone.

"Suck it and see!" A quotation from Dr Brian McArdle (Department of Statistics, University of Auckland), frequently given in response to a statistically oriented question.

# 1 Introduction

Multi-Dimensional Scaling (MDS), trees and planar graphs are three different and popular ways of presenting complex multidimensional data. Sometimes there is a clear reason for preferring one over the others. For example, trees are a natural fit with evolutionary theory for a single locus or recombinational unit (sometimes equated with a gene) and often describe well the pattern of descent with modification of whole genomes (e.g., Swofford et al. 1996). Multi-Dimensional Scaling (MDS, e.g., Edwards and Oman 2003) has special appeal in genetics when gene flow is predominantly diminished by geographic distance, that is, an isolation by distance model. Planar graphs, such as Neighbor nets, are a generalization of trees that aim to capture some of the potentially contradictory information due to, for example, gene flow between specific populations that were previously separated in a tree-like manner (e.g., Huson and Bryant 2006).

Unfortunately, at present, in fields such as genetics, the choice of which model to present and how to compare them is somewhat arbitrary as are estimates of the support for different aspects of the model (e.g., recent articles Nature and Science Magazine such as Tishkoff et al. 2009, The HUGO Pan-Asian SNP Consortium, et al. 2009 and Rasmussen et al. 2010). Here we address the first problem, that of comparison, in a likelihood framework with particular emphasis on flexi-Weighted Least Squares (fWLS) and the g%SD measure as an intuitive guide to model fit (Waddell et al. 2007, Waddell and Azad 2009). To address which parts of a tree or planar diagram are robust and which are not, we use residual resampling methods like those presented in Waddell and Azad (2009). We expand the use of fWLS to consider both polynomial (for all methods) and exponentially weighted models (MDS and trees) to assess which type of weights and which parameter settings best describe the residuals.

The importance of residual resampling is most realized when there remains significant systematic error in the data (e.g., Hendy and Penny 1993, Waddell 1995). One way of attempting to represent this unseen structure/systematic error is a Neighbor-net (Bryant and Moulton 2004). At present, support for both trees and such networks are gauged using a site-specific bootstrap (e.g., Felsenstein 1984, Penny and Hendy 1984, Beran 1988) as implemented in programs such as PAUP* (Swofford 2000) and SplitsTree4 (Huson and Bryant 2006). However, these predominantly capture sampling error under the assumption that all characters are independent (e.g., Waddell et al. 1994). Thus their use with Neighbor-net prior to a final interpretation of the data can be somewhat analogous to "the pot calling the kettle black;" a Neighbor-net network may be suggested when there seems to be systematic error and it may look both informative and robust, but its splits may be no more unambiguously identified than those in a tree when the total residual error is considered.

We illustrate these methods with a genetic data set based on ~500,000 single nucleotide polymorphisms looking at the genetic relationships of Abraham's Children, that is Jewish and Arabic semitic people living in Europe and the Middle East (Atzmon et al. 2010). The earlier analyses presents a wide-ranging analysis of the genetic origins of Jewish populations using currently popular techniques. Here we show how a more unified framework for MDS, trees and Neighbor nets offers useful insights on these populations and the challenges of fitting models to population genetic data.

# 2 Materials and Methods

The original data were a collection of single nucleotide polymorphisms. Jewish individuals were genotyped by Atzmon et al. (2010) while non-Jewish populations were genotyped by (Rosenberg et al. 2002). After screening out related individuals, poorly called (<95%) SNP's, and keeping overlapping SNP's between the two data sets, there were a total of 164,894 SNP's remaining (Atzmon et al. 2010). The $F_{st}$ between populations was calculated and a distance estimated according to Weir and Cockerham (1984). This distance matrix appears in table 1 of Atzmon et al. (2010) to only one or two significant places. These distances to eight decimal places were kindly sent by one of the authors (Li Hao).

To calculate fWLS squares trees (an extension of the least squares trees of Cavalli-Sforza and Edwards 1967 and Fitch and Margoliash 1967) and Neighbor-joining trees (NJ, Saitou and Nei 1987), three different programs were used. FITCH (Felsenstein 1989) and PAUP*4_a114 (Swofford 2000) were used to search for least squares fitted trees that had weights of the form distance$^P$, where P is a free parameter (we call these polynomial weights or the standard model, e.g. Waddell, Tan and Khan 2010). C and Perl scripts were written to run these searches at a range of values of P from -4 to 6. For weights of the form exp(P'distance), we modified the code of FITCH and recompiled the program (we call these exponential weights or a multiplicative model, e.g. Waddell, Tan and Khan 2010). Edge lengths were forced to be non-negative when searching for the best tree (e.g., Swofford et al. 1996). For the calculation of iteratively weighted least squares (for either of these models) a script was written in Matlab using its matrix operation functions. Since the effect of these weights is dependent upon the scale of the distance, we would select the range or ratio of the weights for the multiplicative model to approximately match that of the standard model.

Herein, the g%SD values are calculated with the root mean square error term using $1/(N-k)$ where $N$ is the number of non-trivial distances and $k$ is the number of parameters fitted to the model (Waddell, Tan and Khan, 2010). Residual resampling used our own C program and the consensus trees were estimated using CONSENSE (Felsenstein 1989).

Multidimensional scaling was performed with extensively modified versions of the R routine discussed in Edwards and Oman (2003). These implemented fWLS according to both the polynomial and exponential weighting system. The optimization landscape for minimizing the least squares function can be bumpy, so the starting points for both 2 and 3 dimensional MDS was chosen at random and the whole optimization repeated 1000 times at each value of P or P'. Even then there would sometimes still be some lack of smoothness between adjacent values, so the best solutions for each random replicate were kept and tried for all values of P and P' to give the minima for the final figures presented below.

The planar graphs were estimated using SplitsTree4 (Huson and Bryant 2006). The graph was selected using Neighbor-net and the sum of squares, fit, and residual resampling replicates were calculated with our own C programs using the three values of polynomial weights available in the program for estimating edge lengths, namely, P = 0, 1 or 2. Excel was used for miscellaneous calculations and plots. To estimate balanced minimum evolution (BME, Pauplin 2000) trees with SPR (e.g., Hordijk and Gascuel 2005) and TBR searches we are grateful to Olivier Gascuel and Vincent Lefort for an unpublished version of fastME (Desper and Gascuel 2002).

## 3 Results

The results follow the order of inferring and assessing the robustness of fWLS trees for the genetic distances between Abraham's children (according to the old testament of the bible, these are the Bedouin plus related peoples and the now widely dispersed Jewish groups), a selection of Europeans, and an combined African genetic profile as the outgroup. After this, fWLS MDS models are fitted, followed by fitting a Neighbor Net and assessing its robustness using residual resampling, and finally a comparison of the fit of the different models, including information criteria such as AIC and BIC.

### 3.1 Polynomially and Exponentially flexi-Weighted Least Squares Trees

The first task is to find the optimal P and P' values for fWLS trees with polynomial and multiplicative weights with results shown in figure 1a. Since the effect of different values of P and P' vary widely and independently with the scale of the distances, the x-axis reflects the range of weights divided by the mean weight. It is clear the two models are behaving very similarly on this data. For an alternative x-axis consider figure 1b, which uses the log to base 10 of the ratio of the weight of the largest distance versus that of the smallest distance. On figure 1b it is seen that the exponential weights do best, with a g%SD of 10.353, when the ratio of the weight of the largest to the smallest distance is close to $10^1$ or 10. The polynomial weights achieve a slightly better fit, with a minimum g%SD of 9.378, when the ratio is around $10^3$ or 1000.

Notice the jaggedness of the blue line (polynomial weights) in figure 1b. This occurred using the PHYLIP program FITCH with P in the range 5 to 6, thus fairly extreme weights. This was due to the extreme weights creating a "bumpier" tree landscape and the relatively simple nearest Neighbor interchange algorithm of FITCH being unable to reliably find the best tree despite twenty random addition starts. The green line shows that this issue is dealt with in PAUP* through the use of more general SPR and TBR tree rearrangements. Such bumpiness in the fit plotted against the variability of weights is seen also using MDS and there too, it tends to first show up with the more extreme weights. We also observed some bumpiness in the fit of the multiplicative, but not the polynomial weights, at extreme negative values (log to base 10 ratios of -5 to -10 and g%SD of ~ 100, thus off our charts). This however, might also have been a second local minima, as seen in the yeast data set with the same model (Waddell, Tan and Khan 2010), although here clearly at much worse fits than the global minimum at around P' = 2.2.

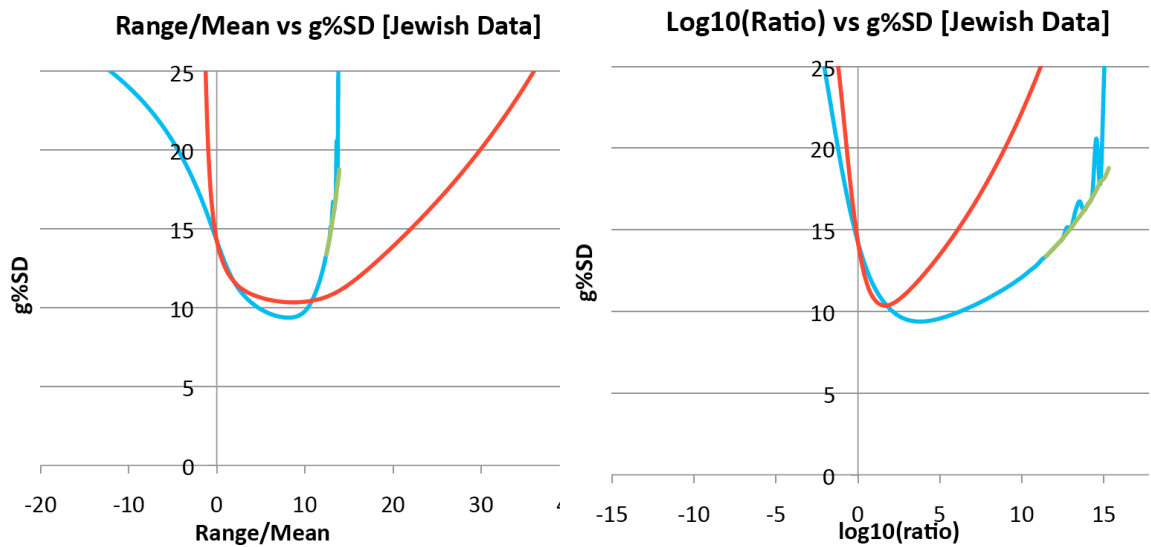

Figure 1 (a). A plot of the fit of the standard model (blue) and of the multiplicative model (red) versus the range of weights divided by their mean for fWLS trees fitted to the Fst genetic distances of Atzmon et al. (2010). (b) The same, but with the x-axis rescaled to represent the log to base 10 of the ratio of the weight of the largest distance divided by that of the smallest distance. The minima are at P = 1.5 and g%SD = 9.378 for the standard model, and P'= 2.2 and 10.353 for the multiplicative model. By contrast, the value at P = P' = 0 is g%SD = 14.236. The green line indicates part of the curve reestimated using PAUP* with a more general tree search strategy.

For both models the fit of a tree to this data is not great, so it is important to have a robust estimate of the stability of the reconstructed tree(s). Both models optimal trees were the same with relatively minor differences in the reconstructed edge lengths. The polynomial weighted tree is shown in figure 2. The general form of the tree is very close to that of figure 2c of Atzmon et al. (2010) with the exception of the position of the Caucus population of the Adygei. Features include the more pure semitic populations, such as Bedouin, Palestinians and Druze, at the root of the tree, followed by non-monophyletic groups of dispersed Jewish populations, including

Iranian, Iraqi, Syrian, Turkish, Greek, Italian and Ashkenazi Jews, followed by European groups. The intermediate position of the Jewish populations on the tree could be due to some members of this group, including the Italian and Ashkenazi groups, being effectively hybrid populations, comprised of roughly half semitic and half European genes, and dragging along with them the other Jewish groups with more Middle Eastern genes.

The residual resampling confidence in edges of the tree (Waddell and Azad 2009) as shown in figure 2, is markedly lower than that shown in the earlier work of Atzmon et al. (2010) using NJ and standard character bootstrapping. This is consistent with the ~10% error on the average distance when fitted to a tree model. None the less, the more strongly supported parts of the tree do highlight the general conclusions of Atzmon et al. (2010) that the Jewish populations are intermediate between those of Europe and those of the Middle Eastern semitic peoples. This type of placement remains consistent with a hybrid European origin of the genes of at least some populations, which are then possibly dragging other Jewish groups along. This hypothesis can be tested by sequentially removing the Jewish populations that show the most admixture with Europeans (in the order of Ashkenazi, Italian, Greek, Turk, Syrian, Iranian and Iraqi Jews according to the Structure analyses in Atzmon et al. 2010). Partly consistent with this, the removal of the European Jews (Ashkenazi, Italian and Greek) sees a major rearrangement in the Neighbor joining tree, with the Adygei population moving several nodes to become sister to the Europeans, as it is in the fWLS trees. However, it is not clear that the removal of any particular Jewish groups sees the remainder move sister to any of the other Middle Eastern groups, so there is no clear support for the hypothesis that some Jewish groups are being dragged along by others.

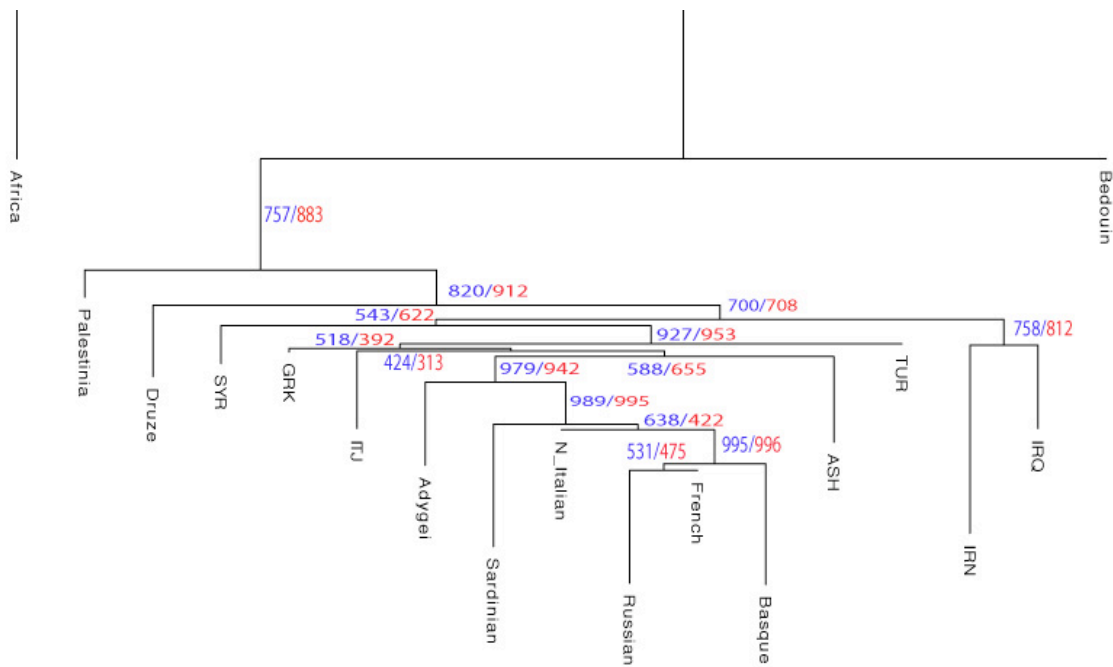

Figure 2. The optimal tree for fWLS for both polynomial and exponential weights. The edge lengths shown are those of the slightly better fitting polynomial weights. The blue values are the consensus tree values for 1000 residual resampling following the standard model, while the red values are those of the multiplicative model. Note, that the most strongly supported groups with both models are Europeans, including European Jews, Europeans excluding European Jews, Europeans excluding the Adygei, and the more northerly Russian, French and Basque group. There are claims that the Druze are a Jewish group that converted their religion (Lev, 2010), and this tree does not exclude that possibility, showing the Druze to partition closer to the Jewish/European populations than other Middle Eastern groups, including the Palestinians, with considerable support.

As seen in other recent work (e.g., Waddell, Azad and Khan 2010), the fit of BME type models can be poor and in this case the NJ tree has a g%SD of > 30 (table 1). This translates to very weak support for all internal edges in the residual resampling consensus tree, which is (((((((IRQ:1000,IRN:1000):271,((Palestinia:1000,(Bedouin:1000,Africa:1000):825):763,Druze:1000):275):51,(SYR:1000,GRK:1000):70):20,(ITJ:1000,ASH:1000.):100):55,TUR:1000):303,(((French:1000,Russian:1000):645,(Sardinian:1000,Basque:1000):465):277, N_Italian:1000):470):1000,Adygei:1000). In many ways this is a better comparison with the tree in Atzmon et al. (2010) showing that the bootstrap values they obtained for NJ using a standard bootstrap captures only a small fraction of the inherent error in the fit of data to NJ tree model. The residual resampling NJ and BME consensus trees are weak enough to discourage any but the most general inferences.

**3.2 fWLS MDS**

In this section we show implement flexi-Weighted Least Squares for multidimensional scaling (MDS). This includes special cases such as unweighted least squares MDS as well as Sammon's Stress Function which in our case is P = 1 (Sammon 1969, Waddell, et al. 2007). We implement both polynomial and exponential weighted models. A special challenge with MDS is that it is potentially an NP complete problem requiring a vast exploration of the starting parameter space to ensure that numerical optimization climbs to the top of a global and not a local optimum. We deal with this by starting the quasi-Newton optimization from 100 random starting points for every step in the parameter P or P', we run 100 steps and keep the best solution at each value. This can still lead to a bumpy curve, particularly for the more extreme weights which, as seen with trees, seem to create a rougher search landscape. To smooth this curve, we also run every optimal solution found for every other value of P or P', in case it is a better solution than the random starts at a particular parameter value. We also have an option to start from the PCA coordinate solution (Edwards and Oman 2003), but on average, this does more poorly than the random replicates. Given that the optimal solution may not have been found, we recommend inspection of the plots of the fit versus the weights to assess whether it is likely a near optimal solution has been found near the optimal weighting function. For the Jewish data the results are shown in figure 3. In this case the curve was smooth across the range, of which only the minimum is shown in the figure.

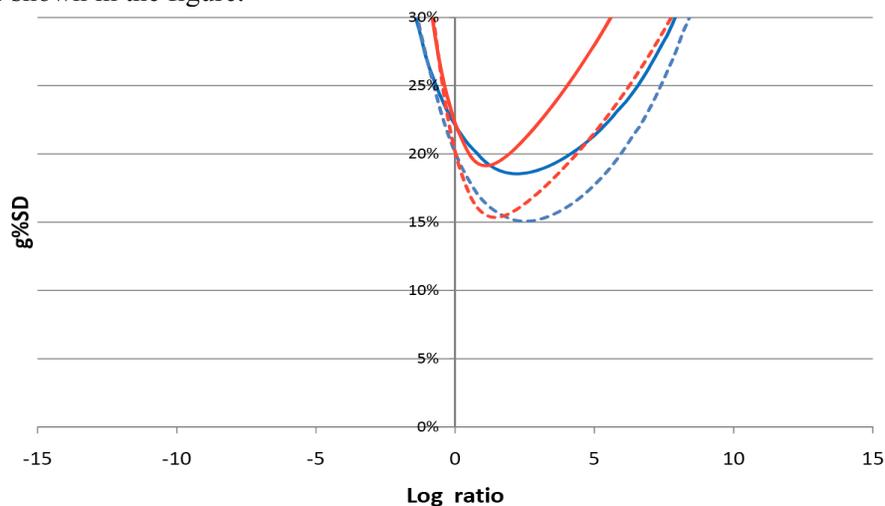

Figure 3. Fit of the MDS plots. The *x*-axis is the log to base 10 of the ratio of the weight on the largest distance divided by that of the weight on the smallest distance. The red lines are polynomial weights and the blue lines are exponential weights, while solid is 2D and dashed are 3D.

Overall the polynomial weights models were slightly better than the exponential weights model, mimicking the result already seen with the trees in figure 1. The 3D models improve fit by ~3.5%, but at the cost of another t-1 fitted parameters. Whether this is a reasonable trade off is assessed in section 3.4.

The best fitting two dimensional (2D) MDS model had polynomial weights with P = 0.9 and g%SD of 18.55. This value may be directly compared to the fit obtained with the trees, since 2D MDS optimizes 2 free parameters (the x and y coordinate) per taxon, minus the invariant point of origin (2 parameters) and a single other fixed y-value of zero. These remove isomorphic solutions due to rotation and reflection and leave exactly $2t - 3$ parameters to optimize, which is exactly the same as a binary tree of $t$ taxa. In this regard, the fit of MDS, while reasonable by the standards suggested by Sammon (1969) is markedly inferior to a tree for this data set. None the less, it is worth examining, as the tree itself is not a perfect fit, and MDS may allow some useful insight.

The 2D polynomial plot show in figure 4 are remarkably map like. Most of the groups fall into the relative positions expected if they were plotted on a geographical map. The great distance to Africa and the Iranian Jews being pushed to the east are expected to be largely reflections of a strong out of Africa bottleneck and a protracted period of in breeding and genetic drift on the Iranian Jewish population. The possibility of genetic drift in the Sardinians is also suggested by their being pushed to the west in the MDS diagram and having a long external edge on the tree. What might be most unexpected from the tree and perhaps most insightful on this diagram is the relocation of the Adygei and the Ashkenazi Jews close together. The location of the Adygei is a compromise between their position on the fWLS and the NJ trees, being halfway between some Europeans and some Middle Easterners. The Structure analyses in Atzmon et al. (2010) also offers the insight of their having a mixed European, Middle Eastern, southern and east Asian ancestry, roughly in that order. The close association with the Ashkenazi Jews is a surprise then since their Structure profile, while not dissimilar to that of the Adygei, looks nearly identical to that of the other European Jewish populations. Whether this a signal due to mixing with Slavic population in the last 1000 years is unclear, but they have moved away from the Sephardic Jews in Greece and Turkey, not directly towards the Russian's but towards the Adygei. The Druze no longer appear specifically connected to Jewish groups, but rather, appear as thought they might have been pulled more by a southern European component, which is also consistent with the Structure analysis in Atzmon et al. (2010).

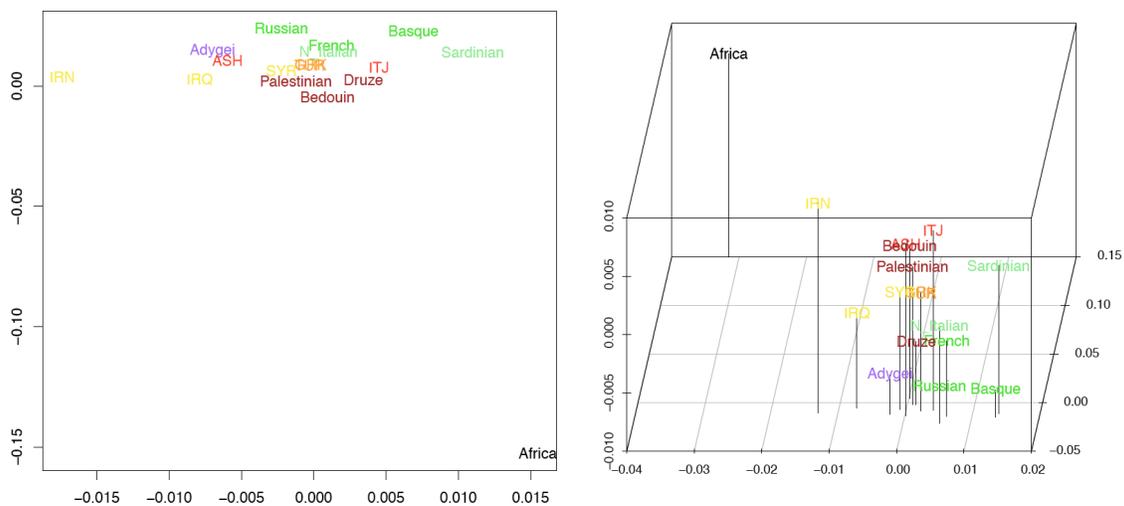

Figure 4 (a). The optimal 2D MDS diagram with polynomial weights, P = 0.9 and g%SD = 18.55. If this is reflected about the vertical it closely approximates the geographic map. The labels record the different populations, with black being the African outgroup, dark brown being the Middle Eastern populations, yellow being Middle Eastern Jewish groups, orange being South Eastern European Jewish populations, red being more central European Jewish groups, light green southern Europeans, dark green more northern Europeans, and purple the Caucus based Adygei. (b) The 3D MDS diagram with polynomial weights, P = 1and g%SD = 15.07.

The 3D plot is somewhat similar in general form to the 2D plot, but offers surprises particularly in terms of the relationships of Jewish populations. All Jewish populations except those in Iran and Iraq appear close to the Palestinians and Bedouin, including the Ashkenazi, that show not special affinity to the Adygei at all. In contrast, it is now the Druze that at first glance appear to be an atypical Middle Eastern group, appearing now to be close to the Europeans.

**3.3 Neighbor Net**

Figure 5 shows the Neighbor net network reported by Splits Trees 4 fro the genetic distances. The program itself estimates the edge lengths after the network is constructed and it can do this at three powers of P (0, 1 and 2) with polynomial weights. The program seems to give a confidence measure for each split in the network, but this is currently a bug (D. Bryant, pers comm.) The network shows appreciable boxiness amongst the European populations and with a non-tree signal pulling Palestinians towards the African outgroup.

To calculate the residual sum of squares we implemented C code to read the network output and compare the observed and the expected split distances. The variance is estimated using the term $N - k$, where $N$ is the number of unique distances and $k$ is the number of splits in the network (which can vary up to nearly order $t^2$, which is a lot more than the order $t$ for splits in a tree, where $t$ is the number of taxa). We then produce 1000 replicates of the input distance matrix by randomly adding independent standardized normal deviates of the observed variance back to the network distances. The Neighbor Net is then built from scratch 1000 times and the proportion of times a split in the original network is contained in the 1000 replicates is counted (irrespective of its weight).

The results of residual resampling are surprising. In order to graphically show these on a complicated network, the color scheme of Waddell et al. (1999) is extended, with extra divisions showing the very highest and lowest support. Despite the tree showing many strongly supported internal divisions and despite many of these appearing in the network, they receive mostly weak network support upon resampling. It is also revealing that the more heavily weighted splits (e.g., the separation of Adygei and Russians) do not necessarily fare better upon residual resampling. In summary, there is no part of the network that consistently contradicts a tree-like interpretation of the data based upon residual resampling, but there is relatively weak structure in resolving any of the internal splits.

Model selection is very sensitive to the effective number of alternative parameters available to choose from (e.g., Chen and Chen 2008). The effective alternatives in a tree are highly restrained by each strongly supported internal edge. The parameter space of planar networks is much more general and presents a much wider choice of parameters for inclusion. It may be this fact that is being most strongly revealed by residual resampling. In this light, one interpretation of the result here is to not to be overly wedded to any specific alternative to a tree, unless the network shows stable non-tree parts, which they do not do in this case. Given the complexity of the parameter space, the question of the consistency of networks is considerably more complicated than that of tree building methods where a standard BIC criterion added to maximum likelihood can ensure consistency in choosing the correct number of edges if some edges in the generating tree are truly zero (e.g., Waddell 1998).

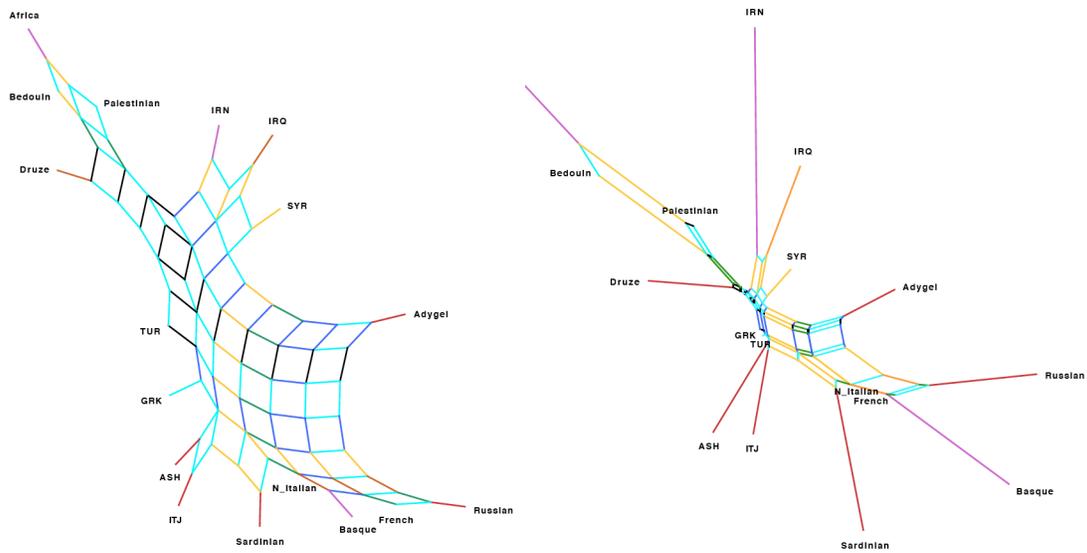

Figure 3 (a) The unweighted and (b) weighted Neighbor Net networks. The colors indicate how often a split in the Neighbor Net was recovered after residual resampling (Waddell and Azad 2009) was applied, and the network repeatedly rebuilt. Black indicates 0-5%, dark blue indicates 5-10%, light blue 10-33%, green 33-50%, yellow 50 to 80%, orange 80 to 90%, red 90 to 95% and purple 95 to 100%. Note, even the most robust splits in the trees are now ambiguous.

Exploration of the network by including and excluding populations shows that the exclusion of the Adygei results in the network looking surprisingly tree-like, and the fit statistic shows a marked but not a massive improvement. The shape of the network is now very close to that of the NJ or fWLS tree with the Adygei removed. Interestingly, at this point a small split does appear separating the Ashkenazi Jews, French and Russians from the rest. It is not clear if this offers any evidence for a Slavic influence on the Ashkenazi, and could perhaps be a north central European signal from their purported residence in the Rhine valley in the period before 1000AD (Atzmon et al. 2010).

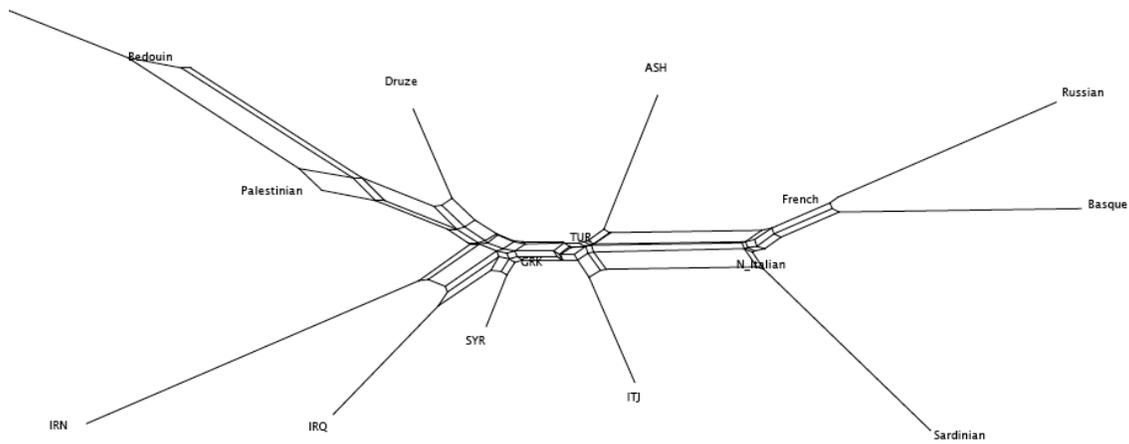

Figure 4 Weighted Neighbor Net network after the removal of the Caucus Adygei populations. The diagram has become dramatically more tree-like, but and the fit has improved somewhat with the proportion of the variance explained going from 95.59 to 96.806 at P = 1. The truncated external edge at the top right goes to the African outgroup.

### 3.4 5 Comparison of fit for the different model types

In this section the fit of the different models for trees, MDS and Neighbor Net are compared and contrasted using a variety of fit statistics. These include intuitive measures such as the percent standard deviation (%SD) and the geometric mean percent standard deviation (g%SD). The former is not optimized directly here, but simply reported when the g%SD was minimized. The latter has the property of being monotonic with the likelihood of the data and of being more comparable across data sets. The log likelihood of the data is also presented, for polynomial (Sanjuán and Wróbel 2005, Waddell et al. 2007) and exponential fWLS models (Waddell, Tan and Khan 2010). From the log likelihood, it is possible to use model selection procedures such as the Akaike Information Criterion (AIC, Akaike 1974), AICc or AIC adjusted for small samples (Burnham and Anderson 2004), and the Bayesian Information Criterion (BIC, Swartz 1978). Because we are using an empirical estimate of the variance, the fit statistics here are behaving more like quasi-AIC, QAIC, QAICc, etc.

The commonly used tree building BME methods, NJ and a TBR BME search, both yielded poor fit to the data (the full log likelihood was not evaluated, but the g%SD is very poor). This is consistent with their fit to data sets seen elsewhere (e.g., Waddell and Azad 2009, Waddell, Azad and Khan 2010). The ordinary least squares model constrained to non-negative edge lengths is a marked improvement, but the fWLS models with edge lengths constrained to be positive do markedly better by log likelihood, g%SD, AIC, AICc and BIC.

The various multidimensional scaling models do not do especially well in comparison to fWLS and the 3D models offer little over the 2D models in the adjusted information criteria (AICc and BIC). The %SD measure on the exponentially weighted MDS diagrams heads off to infinity (indicated with the word huge in table 1 and > 100,000), because the arithmetic mean of the weights (as opposed to the geometric mean) blows up.

Nearly all the fit measures favor the Neighbor Net over all the alternatives. However, the Bayesian Information criterion favors the fWLS polynomial tree over the Neighbor Net, with the exponentially weighted fWLS tree in third place, followed by a marked gap back to the other models.

Table 1. Comparison of different types of model fitted to genetic distances by their sum of squares, optimal parameter values, relative log likelihoods, number of fitted parameters ($k$), mean percent standard deviation (using the term $1/(N-k)$), geometric percent standard deviation (using the term $1/(N-k)$), then the Akaike, small sample adjusted Akaike and Bayesian Information Criteria. For all measures, except lnL, lower is better.

| Model   | SS      | P/P' | lnL    | k  | %SD  | g%SD | AIC       | AICc      | BIC       |
|---------|---------|------|--------|----|------|------|-----------|-----------|-----------|
| NJ      | 0.00355 | na   | < 717  | 31 | -    | 30.2 | horrible  | horrible  | horrible  |
| BME     | 0.00216 | na   | < 751  | 31 | -    | 20.5 | poor      | poor      | poor      |
| OLS+    | 0.00054 | 0.0  | 845.7  | 31 | 6.5  | 14.3 | -1629.4   | -1610.3   | -693.4    |
| fWLS-P  | 0.11670 | 1.5  | 902.1  | 31 | 7.9  | 9.4  | -1742.3   | -1723.2   | **-749.8** |
| fWLS-P' | 0.01451 | 2.2  | 888.7  | 31 | 5.3  | 10.4 | -1715.4   | -1696.3   | -736.4    |
| MDS-P   | 0.03814 | 0.9  | 809.4  | 31 | 11.4 | 18.5 | -1556.7   | -1537.7   | -657.1    |
| MDS-P'  | 0.00063 | 14.4 | 805.1  | 31 | huge | 19.1 | -1548.2   | -1529.1   | -652.8    |
| MDS3-P  | 0.03229 | 1.0  | 848.8  | 47 | 8.8  | 15.1 | -1603.7   | -1552.4   | -617.9    |
| MDS3-P' | 0.00030 | 19.2 | 846.3  | 47 | huge | 15.4 | -1598.6   | -1547.3   | -615.4    |
| NN-P    | 0.69422 | 2.0  | **921.5** | 36 | 8.3  | **8.3** | **-1771.1** | **-1744.2** | -744.7    |

## 4 Discussion

In summary, the comparison of trees, MDS and Neighbor Nets in a common framework allows for a more integrated interpretation of complex data sets such as human historical population genetics. In this particular case, we see that certain types of tree, in particular fWLS trees, fit the data better than MDS or commonly used tree building approaches such as NJ. In comparison Neighbor Net also fits about as well as fWLS trees. Indeed it fits better by many criteria than fWLS except for the BIC criterion. Given the complexity of Neighbor Net models in terms of the "parameter choice space" which grows as order $t^2$ compared to order $t$ for trees, it may be necessary to impose a heavier type of BIC penalty upon them in future (e.g., Chen and Chen 2008).

Since the fit of all these models left quite a lot to be desired (e.g., g%SD was never better than 8%), the use of residual resampling gives much more robust estimates of the stability of models. In comparison, a standard bootstrap on the SNP's which number over 100,000 is likely to reveal only a very small amount of multinomial based sampling error (we expect less than 2%SD in this cases). This turns out to be especially important in the case of the Neighbor Net, where more realistic estimates of the residual error and the complex parameter space, see the identification of particular splits in the data (i.e., the favored model) become very weak. This cautions us not to over interpret the Neighbor Net network inferences.

In terms of the history and genetics of Jewish people in particular, these analyses leave a few tantalizing hints. While the trees do not show it, the MDS and Neighbor Net do suggest the possibility of a specific genetic association between the Ashkenazi Jews alone with East Europeans (in this analysis Russians and perhaps the Adygei). This is likely to remain a topic of some interest, not least for eugenicists who are intrigued by the possible genetic basis for the high IQ scores of the Ashkenazi (e.g., Cochran et al. 2006). In terms of where and how many Jewish populations picked up their considerable European genetic component, this remains unclear, since even the Jewish populations of middle earth (Syria, Iran and Iraq) that were separated in the early Diaspora, cluster closer to Europeans than other Levant populations such as the Druze in the trees and Neighbor Net. The analyses also leave open the possibility that the Druze are a Jewish group that converted religion. With their complicated genetic history, further study of Jewish groups is likely to be a real bonus for all groups of people that want a more detailed deep genealogical/genetic history.

Using the type of framework outlined here, in future, it will be possible to combine MDS and trees into one framework. An additional weighting parameter will determine how much of the expected distances are contributed by the MDS versus the tree component of the model. Such a model may be useful when an initially distinct tree-like structure begins to be overwritten with a separation by distance genetic model. One question here is whether such models are identifiable given just the genetic distances, or whether they require the full haplotype data to offer robust estimates.

## Acknowledgements

This work was supported by NIH grant 5R01LM008626 to PJW. Thanks to David Bryant, Joe Felsenstein, Olivier Gascuel, and Hiro Kishino for helpful discussions. Thanks to Li Hao for sending the distance matrix from table 1 of Atzmon et al. (2010) to eight decimal places of precision.

## Author contributions

PJW originated the research, developed methods, gathered data, ran analyses, interpreted analyses, prepared figures and wrote the manuscript. IK XT, and SY implemented methods in C, PERL and R, ran analyses, prepared figures, interpreted analyses and commented on the manuscript.